\documentclass[a4paper,12pt]{article}
\pdfoutput=1

\usepackage[utf8x]{inputenc}
\usepackage{epsfig} 
\usepackage{amssymb}
\usepackage{amsfonts}
\usepackage{amsmath}
\usepackage{mathrsfs}
\usepackage{nicefrac}
\usepackage{color}

\makeatletter \@addtoreset{equation}{section} \makeatother

\begin{document}
\begin{titlepage}
{\hfill     IFUP-TH/2012-07}
\bigskip

\begin{center}
{\huge  {\bf
Singular points in $\mathcal{N}=2$ SQCD.
 } }
\end{center}

\bigskip
\begin{center}
{\large  Simone Giacomelli   \vskip 0.10cm
 }
\end{center}

\begin{center}
{\it   \footnotesize
Scuola Normale Superiore - Pisa ,
 Piazza dei Cavalieri 7, 56126 Pisa, Italy.\\
Istituto Nazionale di Fisica Nucleare -- Sezione di Pisa, \\
     Large Bruno Pontecorvo, 3, Ed. C, 56127 Pisa,  Italy.
   }

\end {center}

\noindent
{\bf Abstract:}
We revisit the study of singular points in $\mathcal{N}=2$ SQCD with classical gauge groups. Using a technique proposed recently 
by Gaiotto, Seiberg and Tachikawa we find that the low-energy physics at the maximally singular point involves two superconformal 
sectors coupled to an infrared free $SU(2)$ gauge group. When one softly breaks extended supersymmetry to $\mathcal{N}=1$ 
adding a mass term for the chiral multiplet in the adjoint representation, a finite number of vacua remain and the theory 
becomes confining. Our analysis allows to identify the low-energy physics at these distinguished points in the moduli space. In 
some cases, which we will describe in detail, two sectors coupled to an infrared free $SU(2)$ gauge group emerge as before. For USp and SO gauge groups one of these
sectors is always free, contrary to the SU case.

\vfill

\end{titlepage}

\tableofcontents

\section{Introduction}

Theories with $\mathcal{N}=2$ supersymmetry represent an interesting theoretical laboratory to study 
nonperturbative effects in four dimensions. The great virtue of these models is that the low energy dynamics is explicitly known and can be
encoded in a family of algebraic curves, as shown for $SU(2)$ gauge theories by Seiberg and Witten in \cite{SWI,SWII} and then for more general gauge groups
in \cite{APSH,ASI} (see also \cite{TT} and references therein).

It was soon realized that superconformal fixed points are ubiquitous in these models
and their study was initiated in \cite{AD} and \cite{ADSW}. In particular, in the latter reference general properties of conformal theories with
$\mathcal{N}=2$ supersymmetry were derived, such as the fact that mass parameters associated to a nonAbelian global symmetry cannot acquire
anomalous dimension. A more systematic analysis was initiated in \cite{EH} and especially \cite{EHIY}, in which the authors classified singular
points in $SU(N)$ SQCD with $N_{f}$ flavors. All these papers are based on the idea that the SW curve in a neighbourhood of the singular
point should exhibit scale invariance. Combining this with the requirement that the SW differential has scaling dimension one fixes
the scaling dimensions of all the chiral operators. This analysis revealed the existence (for any value of the bare mass $m$ of the flavors) of singular submanifolds in the 
moduli space such that the SW curve factorizes
as $y^2=(x+m)^{2r}Q(x)$ ($2r\leq N_f$). The low energy dynamics is described by a nonAbelian $SU(r)$ theory with $N_f$ massless matter fields in the fundamental. 
Only in the case $2r=N_f$ we have an interacting fixed point. Tuning $m$ appropriately one can find points in the moduli space where
the curve becomes more singular. In this latter case the approach of \cite{EHIY} leads to anomalous dimensions for the casimirs of the nonAbelian flavor group,
contrary to the argument given in \cite{ADSW}. 

More recently the situation has been reanalyzed in \cite{GS}, in which the authors show that this problem can be solved if one
allows for the existence of two scale invariant sectors, weakly coupled by a gauge field. This is analogous to Argyres-Seiberg
duality \cite{AS} (and generalization thereof), apart from the fact that the gauge group appearing in the dual description is infrared free. 
This proposal also removes a possible counterexample to the a-theorem \cite{STI} (see also \cite{KS,KA}).

In \cite{EHIY} the authors analyzed singular points in $USp(2N)$ and $SO(N)$ gauge theories as well. The result of their study is that, as long as
the flavors are massive, the singular points are identical to those of $SU(N)$ SQCD. However, when the masses are set to zero the flavor symmetry enhances
and one finds a different class of fixed points. At that time no tools were available to study them but now, with the techniques of \cite{AS} and
the methods developed by Gaiotto in \cite{G} and Tachikawa in \cite{DT}, the problem can be approached. The scope of this note is to make a systematic
analysis of these singular points.

Another fundamental aspect of $\mathcal{N}=2$ gauge theories is that breaking softly extended supersymmetry with a mass term for the chiral superfield in the 
$\mathcal{N}=2$ vector multiplet, the theory becomes confining \cite{SWI}: the moduli space is lifted and a finite number of vacua remain in which 
magnetically charged objects condense realizing the 't Hooft-Mandelstam
mechanism for confinement. The low energy dynamics of these models have been extensively studied (an incomplete list of references is \cite{APS,CKM,APSII,CKMII}).    
It turned out that the relevant vacua are generically characterized by a nonAbelian gauge symmetry in the infrared. When the flavors are massive the
properties of these vacua are ``universal'' and do not depend on the gauge group (for classical gauge groups). In the massless limit this picture does not
change for $SU(N)$ theories, whereas a different phenomenon occurs for $SO$ and $USp$ theories \cite{CKM,CKMII}: only two vacua remain. One is characterized
by the condensation of baryonic-like composite objects and is in a nonAbelian Coulomb phase (this vacuum was identified in \cite{APSII}). The second one
(called Chebyshev point in \cite{CKM,CKMII}) arises from the collision of the other vacua and is in general characterized by a strongly interacting low energy theory, which exhibits conformal
invariance in the $\mathcal{N}=2$ limit. In these sense the study of confinement in the softly broken theory and the analysis of singular points
in the parent $\mathcal{N}=2$ theory are linked. As we will see, the analysis of maximally singular points will allow us to understand the low energy
physics at the Chebyshev point as well. The properties of the theory once the $\mathcal{N}=1$ perturbation has been turned on will be studied in a
separate publication.

The paper is organized as follows: in section 2 we review the argument given in \cite{GS}, which will be the key ingredient of our analysis, and explain
the properties of vacua relevant for the perturbation to $\mathcal{N}=1$. In section 3 we determine the structure of the maximally
singular point and the Chebyshev point in $USp(2N)$ gauge theory with $2n$ flavors. In section 4 we repeat this analysis for $SO$ gauge theories
and we conclude with a discussion in section 5. As a byproduct we will recover many of the infinite coupling dualities proposed recently.

\section{$SU(N)$ SQCD with $2n$ flavors and r-vacua}

We will now sketch the argument presented in \cite{GS}. 
As is well known, the SW curve and differential for $SU(N)$ SQCD with $N_f=2n$ flavors are
$$y^2=P_{N}(x)-4\Lambda^{2N-2n}\prod_{i=1}^{2n}(x+m_i),\quad \lambda=xd\log\frac{P_N-y}{P_N+y},$$ with $P_{N}(x)=x^N-\sum_{k\geq2}u_kx^{N-k}$.
For our purposes, it is convenient to rewrite the curve in the following way (see \cite{GS} for the details):
$$y^2=(x^N-\dots-u_N)(x^N-\dots+(4\Lambda^{N-n}-u_{N-n})x^n-\dots-u_N)-\sum_{k=2}^{2n}c_kx^{2n-k}.$$
Setting all the coulomb branch coordinates $u_i$ and casimirs of the flavor symmetry $c_i$ to zero, we find the maximally singular (EHIY) point, where the SW curve and SW differential become: 
$$y^2=x^{N+n}(x^{N-n}+4\Lambda^{N-n}), \quad \lambda\approx\frac{y}{x^n}dx.$$ 
Requiring that the SW differential has dimension one gives the relation $[y]=1+(n-1)[x]$. If we further impose the scale invariance
of the curve we find the equation $2[y]=(N+n)[x]$. These relations fix the scaling dimensions of $x$ and $y$ and in particular
imply an anomalous dimension for the cubic and higher casimirs of the flavor group ($[c_i]=(2N+i)/(N+1)$). So, when $n$ is at least two, the above analysis
is inconsistent with the general constraints for theories with $\mathcal{N}=2$ superconformal symmetry (e.g. the $c_i$'s should have 
canonical dimension \cite{ADSW}). 

A natural resolution of this inconsistency is to identify subsectors with different scalings of $x$; clearly the $N+n$ colliding branch points will
distribute among the subsectors. The proposal of \cite{GS} is precisely along this line: the authors introduce a particular
scaling limit in which two subsectors emerge: one is a $D_{N-n+2}$ Argyres-Douglas theory \cite{EHIY,CV} (or maximally singular point of $SU(N-n+1)$ gauge
theory with 2 flavors \cite{GS}; see also \cite{BMT,AMT}) and the other can be described as a three punctured 
sphere in the Gaiotto framework \cite{G}, as we will now see.

The first step is to rewrite the curve in a ``6d form'': one defines $t=y/x^{n-1}$ (so the SW differential becomes $\lambda\approx tdx/x$) and writes the curve as
\begin{equation}\label{curva}
\begin{aligned}
t^2=&(x^{N-n+2}-u_1x^{N-n+1}-\dots-u_{N-n+2}-\dots-\frac{u_N}{x^{n-2}})\\
&\times(x^{N-n}-\dots+(4\Lambda^{N-n}-u_{N-n})-\dots-\frac{u_N}{x^n})-\sum_{k=2}^{2n}c_kx^{2-k}.
\end{aligned}
\end{equation}
Notice the presence of $u_1$, which is a parameter proportional to $\sum_i m_i$ and not a coordinate on the Coulomb branch. This will
be important in later sections.
To account for the presence of two sectors, we now introduce two scales $\epsilon_A,\epsilon_B\ll1$. Notice that with the above rescaling the 
condition $[\lambda]=1$ implies $[t]=1$ in both sectors.
In the A sector ($\vert x\vert\sim\epsilon_A$) we will impose $[x]=1$, in order to satisfy the constraint $[c_k]=k$. This leads to
the relation $c_k\sim O(\epsilon_{A}^{k})$. Consider now the term $4\Lambda^{N-n}x^{N-n+2}$. It is 
clearly negligible for $\vert x\vert\sim\epsilon_A$ (with respect to, e.g. $\sum_{k=2}^{2n}c_kx^{2-k}$), and consequently has to appear in the B sector
($\vert x\vert\sim\epsilon_B$). This implies $t^2\sim\epsilon_{B}^{N-n+2}$, and since $t$ has scaling dimension one in both sectors, we
deduce the relation \begin{equation}\epsilon_{A}^{2}=\epsilon_{B}^{N-n+2}.\end{equation}
Interestingly, the above considerations and the requirement that all the coulomb branch coordinates appear in at least one sector necessarily imply
 $$u_{k}\sim O(\epsilon_{B}^{k}),\quad k=1,\dots,N-n+2;\quad u_{k}\sim O(\epsilon_{A}^{k+n-N})\quad k=N-n+2,\dots,N.$$  

Now it is possible to read from (\ref{curva}) the curves for the two subsectors just collecting the leading order terms:
\begin{enumerate}
 \item For $\vert x\vert\sim\epsilon_{A}$ we are left with
\begin{equation}
\begin{aligned}
t^2=-(u_{N-n+2}+\dots+\frac{u_N}{x^{n-2}})(4\Lambda^{N-n}-\frac{u_{N-n+2}}{x^2}-\dots-\frac{u_N}{x^n})-\sum_{k=2}^{2n}c_kx^{2-k}.
\end{aligned} 
\end{equation}
As discussed in \cite{GS}, this is the SW curve (when $n>2$) for the Gaiotto theory obtained compactifying n M5 branes on a sphere with three regular punctures (two are maximal
and one is described by a Young tableau with columns of height $\{n-2,1,1\}$). Its global symmetry group is $SU(2)\times SU(2n)$ and the corresponding
casimirs are $c_i$ and $u_{N-n+2}+c_2/4\Lambda^{N-n}$. 
This is precisely the interacting theory that enters in the S-dual description of $SU(n)$ theory with 2n flavors and its properties have been studied in detail in \cite{CD}.
For $n=3$ this S-duality coincides with Argyres-Seiberg duality and the A sector describes the $E_6$ theory of Minahan and Nemeschansky \cite{MNI}. For $n=2$
the theory becomes free and describes three doublets of $SU(2)$ (the global symmetry is $SU(2)\times SO(6)\simeq SU(4)$).
\item For $\vert x\vert\sim\epsilon_{B}$ we find instead
\begin{equation}
 t^2=4\Lambda^{N-n}(x^{N-n+2}-u_1x^{N-n+1}-\dots-u_{N-n+2})-c_2.
\end{equation}
This is the SW curve for the $D_{N-n+2}$ theory. For $N=n$ this sector is free and describes a doublet of hypermultiplets.
\end{enumerate}
Both theory 1 and theory 2 have a $SU(2)$ flavor symmetry; in our context the diagonal combination has been gauged and the
SW curve for $\epsilon_A<\vert x\vert<\epsilon_B$ describes the tubular region associated to this gauge group 
($t^2=-4\Lambda^{N-n}u_{N-n+2}-c_2$). 

As explained in \cite{GS}, one can now evaluate the beta function for this $SU(2)$ gauge group from the above curve: a closed BPS string
located in the tubular region at constant $\vert x\vert$ describes a W-boson with central charge $a$, whereas a geodesic connecting a branch point at
$\vert x\vert\sim\epsilon_{B}$ and another at $\vert x\vert\sim\epsilon_{A}$ describes a monopole with central charge $a_{D}$
(see \cite{GNM} for a detailed discussion on this point).
\begin{equation*}
\begin{aligned}
 a=&\int_{\vert x\vert=\text{const.}}\lambda=2\pi i\alpha;\quad \alpha^{2}=-4\Lambda^{N-n}u_{N-n+2}-c_2,\\
 a_D=&\int_{\vert x\vert\sim\epsilon_{A}}^{\vert x\vert\sim\epsilon_{B}}\lambda=\alpha\left(\frac{N-n}{N-n+2}\log{\epsilon_{A}}+\text{const.}\right).
\end{aligned} 
\end{equation*}
Using then the relation $\tau=\partial a_{D}/\partial a$ and identifying $\epsilon_{A}$ with the renormalization group scale we obtain
$$\frac{d\tau}{d(\log\epsilon_{A})}=\frac{b_1}{2\pi i}=\frac{1}{2\pi i}\frac{N-n}{N-n+2},$$ where $b_1$ is the one-loop
coefficient of the beta function. We thus learn that this $SU(2)$ group is infrared free. Since the contribution to the beta function from the three punctured
sphere is 3, we can read out the contribution given by the $D_{N-n+2}$ theory: $$b_{D_{N-n+2}}=2\left(1-\frac{1}{N-n+2}\right).$$ Indeed, this matches the
result of \cite{CV} (the calculation can also be performed using the techniques presented in \cite{ST}).

In view of the breaking to $\mathcal{N}=1$, the relevant points in the moduli space are those such that the SW curve factorizes
as \begin{equation}\label{fac}y^2=(x-\alpha)(x-\beta)Q^2(x).\end{equation} As shown in \cite{APS,CKM}, in the case of equal masses $m_i$, these vacua are labelled by an integer $r$ ($0\leq r\leq N_f/2$),
corresponding to the fact that $Q(x)$ factorizes as $Q(x)=(x+m)^r\tilde{Q}(x)$.When the $\mathcal{N}=1$ perturbation is turned on, for $m\gg\Lambda$ the theory is in the Higgs phase, 
whereas in the limit $m\ll\Lambda$ it becomes confining.  For each value of $r$ there are $2N-N_f$ vacua and the low energy
effective theory is characterized by an abelian sector and a nonAbelian one with $U(r)$ gauge group and $N_f$ massless 
matter fields in the fundamental representation. For $r<N_f/2$ the low energy theory is infrared free and admits a lagrangian description.
More interesting is the situation for $r=N_f/2$: in this case the nonAbelian sector of the low energy theory is superconformal and
the scaling dimensions of chiral operators can be determined as in \cite{EHIY} (in this case the casimirs have canonical
dimension and there is no need to introduce two sectors).
For generic values of the mass parameter $m$, this is the whole story. However, if one sets $m$ equal to $$m=\omega^{k}_{2N-N_f}\frac{2N-N_f}{N}\Lambda,$$ 
one can show \cite{noi} \footnote{In \cite{noi} the formula for $m$ is different. The discrepancy is simply due to a different 
convention (which is the one adopted in \cite{CSW}): indicating with $\tilde{m}$ and $\tilde{\Lambda}$ the parameters used in those papers, we have 
$m=\tilde{m}/\sqrt{2}$ and $\Lambda^{2N-N_f}=\sqrt{2}^{N_f}\tilde{\Lambda}^{2N-N_f}$. This point was overlooked in \cite{CH}.}
that some of the r vacua (one for each value of r) collide and the curve becomes more singular. This signals the transition from Higgs to
confinement phase \cite{noi}. In this limit
$\alpha$ or $\beta$ in equation (\ref{fac}) become equal to $-m$ and the SW curve and differential can be approximated as $$y^2\approx(x+m)^{N_f+1},\quad\lambda\approx\frac{y}{x^n}dx.$$
Here we recognize the EHIY point when $N=n+1$. Indeed, as was argued in \cite{EHIY}, the physics of this singular point is that for the EHIY point of $SU(n+1)$
gauge theory with $2n$ flavors. In this case the B sector is given by the $D_3$ theory.

We thus propose that the low-energy physics at this point is described by:
\begin{itemize}
\item An abelian $U(1)^{N-n-1}$ sector with massless hypermultiplets charged under each $U(1)$ factor.
\item The $D_3$ theory (B sector).
\item The scale invariant theory entering in the S-dual description of $SU(n)$ SQCD with $2n$ flavors (A sector).
\item An infrared free $SU(2)$ gauge multiplet coupled to sectors A and B.
\end{itemize}
This is identical to the proposal made in \cite{GS}, apart from the fact that the abelian sector includes hypermultiplets charged
under the various $U(1)$ factors (one for each $U(1)$). This comes from the requirement that the point we are discussing is not lifted
by the $\mathcal{N}=1$ perturbation \cite{APS,CKM}. This is not the case for the EHIY point discussed in \cite{GS} (apart from
the case $N=n+1$).

\section{$USp(2N)$ SQCD with $2n$ flavors}

Let us turn to $\mathcal{N}=2$ gauge theories with $USp$ gauge group and $N_f$ hypermultiplets in the fundamental representation (we
consider only the equal mass case as before). If the bare mass $m$ for the matter fields is different from zero, the flavour symmetry
is $U(N_f)$ and, as we said in the introduction, one recovers the results found in the previous section; in particular the vacua surviving
the $\mathcal{N}=1$ perturbation have exactly the same structure as the r-vacua of $SU(N)$ SQCD (see \cite{EHIY,ASI,ASII,CKM}) and all the superconformal
points are analogous to those described in the previous section. 

More interesting is the case of massless matter fields: the first main difference is the flavor symmetry, which
is enhanced to $SO(2N_f)$. Moreover, all the r vacua merge into a single superconformal point in this limit \cite{CKM} (we will refer
to it as the Chebyshev point from now on, because its location in the Coulomb branch is determined by Chebyshev polynomials 
\cite{CKM}). As the symmetry enhancement suggests, this fixed point is different from those we have seen so far. 
The purpose of this section is to study the superconformal points of massless $USp(2N)$ SQCD with $N_f=2n$.

The SW curve and SW differential for this model are \cite{ASI}:
\begin{equation}\label{USP}
xy^2=[xP_{N}(x)+2\Lambda^{2N-2n+2}\prod_im_i]^{2}-4\Lambda^{4N-4n+4}\prod_i(x-m_{i}^{2}), 
\end{equation}
\begin{equation}\label{diff}
 \lambda=\frac{\sqrt{x}}{2\pi i}d\log\left(\frac{xP_{N}(x)+2\Lambda^{2N-2n+2}\prod_im_i-\sqrt{x}y}{xP_{N}(x)+2\Lambda^{2N-2n+2}\prod_im_i+\sqrt{x}y}\right).
\end{equation}
where $P_{N}(x)=x^{N}-u_1x^{N-1}-\dots-u_N$ and $m_i$ are the masses for the flavors. Note that in the $SU(N)$ case $u_1$ was a parameter 
while in this case is a coordinate on the Coulomb branch. We can now rewrite the curve as 
$$xy^2=(x^{N+1}-u_1x^{N}-\dots-u_Nx+\tilde{c}_{2n})^{2}-4\Lambda^{4N-4n+4}x^{2n}-\sum_{i}c_{2i}x^{2n-i},$$ where $c_{2i},\tilde{c}_{2n}$
are the $SO(4N)$ casimirs. We can further rewrite it as
$$xy^2=(x^{N+1}\dots-u_{N-n+1}x^{n}\dots+\tilde{c}_{2n})(x^{N+1}\dots-(u_{N-n+1}-4\Lambda^{2N-2n+2})x^{n}\dots+
\tilde{c}_{2n})\\
-\sum_{i}c_{2i}x^{2n-i},$$ where we just redefined $u_{N-n+1}$. If we set to zero all $c_i$ and $u_k$, we find the maximally 
singular point, where the curve and differential become
\begin{equation}\label{scft}
y^2=x^{N+n}(x^{N-n+1}+4\Lambda^{2N-2n+2}),\quad\lambda=\frac{y}{x^{n}}dx. 
\end{equation}
We thus come across the same problem found in the previous section: imposing $[\lambda]=1$ and $[y]=\frac{N+n}{2}[x]$ leads to anomalous
dimensions for the nonAbelian casimirs ($c_{2i}=2+(i-1)[x]=2\frac{N-n+1+i}{N-n+2}$). In order to determine the structure of this infrared
fixed point, we can adopt the technique seen before and introduce two different sectors. 

It is now convenient to define $t=y/x^{n-1}$ and rewrite the curve as
\begin{equation}
t^2=(x^{N+2-n}\dots+\frac{\tilde{c}_{2n}}{x^{n-1}})(x^{N+1-n}\dots+\frac{\tilde{c}_{2n}}{x^{n}})-\sum_{i}c_{2i}x^{2-i}. 
\end{equation}
The constraint $[c_{2i}]=2i$ can be satisfied introducing the scale $\epsilon_{A}$ and setting $c_{2i}\sim O(\epsilon_{A}^{2i})$, 
$\vert x\vert\sim\epsilon_{A}^{2}$. This gives $t\sim\epsilon_A$. A second sector emerges as we introduce the scale $\epsilon_{B}$ and set $\vert x\vert\sim\epsilon_{B}^2$. 
The same reasoning adopted in the previous section leads to the relation $t^2\sim x^{N+2-n}$, from which we deduce 
$$\epsilon_{B}^{2N+4-2n}=\epsilon_{A}^2.$$
The Coulomb branch coordinates are then scaled to zero as
$$u_{i}\sim O(\epsilon_{B}^{2i})\; i=1,\dots,N-n+2;\quad u_{N-n+2+i}\sim O(\epsilon_{A}^{2+2i})\; i=0,\dots,n-2.$$ Collecting the leading
terms as before we can now determine the SW curves for the two sectors.

For $\vert x\vert\sim\epsilon_{A}^{2}$ the curve becomes
\begin{equation}\label{ea}
t^2=\left(u_{N+2-n}+\dots+\frac{\tilde{c}_{2n}}{x^{n-1}}\right)\left(4\Lambda^{2N+2-2n}-\frac{u_{N-n+2}}{x}+\dots+\frac{\tilde{c}_{2n}}{x^{n}}\right)-\sum_{i}c_{2i}x^{2-i}. 
\end{equation}
It has $2n-2$ branch points.

The remaining $N-n+2$ branch points appear in the second sector, for $\vert x\vert\sim\epsilon_{B}^2$. The curve becomes in this case
\begin{equation}\label{eb}
t^2=4\Lambda^{2N+2-2n}(x^{N+2-n}-\dots-u_{N-n+2})-c_2. 
\end{equation}
Let us analyze these two regions:
 
For $\vert x\vert\sim\epsilon_{B}^2$ we recognize the curve we have seen before: this is the SW curve for the $D_{N-n+2}$ theory. The only difference with respect to the
$SU(N)$ case is that, as we noticed before, $u_1$ is a coordinate on the Coulomb branch in the present context. The flavor symmetry
of this theory is thus just $SU(2)$. Two special cases are $N=n$, when the theory becomes free and describes a doublet of $SU(2)$,
and $N=n-1$, when the curve becomes trivial and describes an ``empty'' theory \cite{AS}.

The curve for the region $\vert x\vert\sim\epsilon_{A}^{2}$ is new; it has $SU(2)\times SO(4n)$ flavor symmetry and can be described as the compactification on a three punctured sphere
of the 6d $(2,0)$ $D_n$ theory \cite{DT}, as we will now see, for $n>2$. For $n=2$ it becomes a free theory. 

The SW curve for this class of theories can be written in the form
\begin{equation}\label{sw}
\lambda^{2N}=\sum_{k}\phi_{2k}(z)\lambda^{2N-2k},\quad \lambda=v\frac{dz}{z}. 
\end{equation}
The theory is specified by the singularities on the Riemann surface, which are labelled by Young tableaux (in the case of regular punctures) 
as in the $A_{N}$ case. From the Young tableaux one can read out the pole structure of the various k-differentials and then determine the Coulomb branch coordinates using Riemann-Roch
theorem \footnote{Contrary to the $A_N$ theory, in which this is the general recipe, the $D_N$ theory has a further 
complication: the coefficients one extracts using Riemann-Roch theorem obey in general non-trivial polynomial relations and one must take 
this into account in order to extract the true coordinates on the Coulomb branch (see \cite{CDT} for a detailed analysis of this issue). 
However, this will not be important in the present case.}. The pole structure at each puncture can be determined as follows \cite{DT,CDII,CDT}:
\begin{itemize}
 \item Take the longest even row in the Young tableau which occurs with odd multiplicity (in our case the row of lenght four) and
remove the last box. Place it at the end of the next available row (such that the result is a Young tableau). Repeat this operation
until it stops (the resulting Young tableau does not contain even rows with odd multiplicity).
\item Number the boxes of the ``corrected'' Young tableau as follows: start with zero in the first box and number the boxes in the
first row with successive integers. When you reach the end of the row, repeat that number in the first box of the following row and continue.
\end{itemize}
The numbers inserted in boxes number $2,4,\dots,2N$ are the orders of the pole of $\phi_{2},\phi_{4},\dots,\phi_{2N}=(\tilde{\phi}_{N})^2$
at the given puncture. The algorithm for $A_{N}$ punctures is obtained just neglecting the first step (odd degree
differentials $\phi_{2k+1}$ do not vanish in the $A_{N}$ case, and the corresponding degree of the pole is the integer contained in
boxes number $2k+1$).

\begin{figure}
\centering{\includegraphics[width=.3\textwidth]{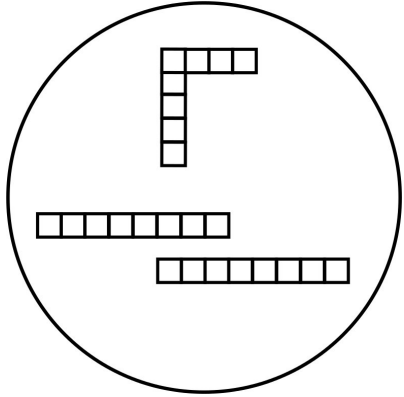}} 
\caption{\label{sdual}\emph{The three punctured sphere that represents the theory entering in the S dual description of $USp(2N)$ with $2N+2$ flavors (in this case $N=3$).}}
\end{figure}

Let us apply the above algorithm to a sphere (depicted in figure\ref{sdual}) with two maximal punctures (labelled by
a Young tableau with a single row of lenght $2n$) and a third one labelled by a Young tableau (always with $2n$ boxes) with a row of lenght 4 and the others
of lenght one (these are all grey punctures in the notation of \cite{DT}). The pole structure at the maximal puncture is $\{1,3,\dots,2n-3;n-1\}$,
whereas the other puncture assigns pole orders $\{1,2,\dots,2;1\}$. The last entry represents the order of the pole of $\tilde{\phi}_{n}$.
The k differentials can thus be written as $$\phi_{2k}=2\frac{u_{2k}z}{(z-1)^2}\left(\frac{dz}{z}\right)^{2k}\; 2k=4,\dots,2n-2;\quad\phi_2=\tilde{\phi}_{n}=0.$$
The SW curve can then be derived just by plugging this result in (\ref{sw}). If we now multiply both sides by $(z-1)^2/v^{2n}$ and
define $y=z-1$, $x=v^2$ we find
$$y^2=2\sum_{k=2}^{n-1}\frac{u_{2k}}{x^k}(y+1)\Longrightarrow \left(y-\sum_{k=2}^{n-1}\frac{u_{2k}}{x^k}\right)^2=\left(\sum_{k=2}^{n-1}\frac{u_{2k}}{x^k}\right)\left(2+\sum_{k=2}^{n-1}\frac{u_{2k}}{x^k}\right).$$
Defining now $(y-\sum_{k=2}^{n-1}\frac{u_{2k}}{x^k})^2=t^2/x^2$ and multiplying both terms by $x^2$ we immediately recognize
(\ref{ea}), with $c_i$ and $u_{N-n+2}$ set to zero. These are the mass parameters associated with the $SO(4n=2N_f)\times SU(2)$
flavor symmetry of the theory. 

Following \cite{GS}, our interpretation is that the infrared physics at the maximally singular point
can be described by the two sectors A ($\vert x\vert\sim\epsilon_A^2$) and B ($\vert x\vert\sim\epsilon_B^2$); both sectors have $SU(2)$ global symmetry and the 
diagonal combination is promoted to a gauge symmetry.

The A sector we have just described (see figure \ref{sdual}) already appeared in \cite{DT} (and for $N=3$ was studied in \cite{CDII}), where it was recognized 
that it enters in the S dual description of the scale invariant $USp(2N)$ theory with $2N+2$ flavors: in the infinite coupling limit the two simple punctures
collide and this three punctured sphere emerges from the collision. Indeed, we can understand this duality using the analysis
of the maximally singular point given above: if we apply the same strategy to the scale invariant case, 
so that the maximally singular point is precisely the origin of the Coulomb branch, we find a B sector which is trivial and thus
the S dual description is given by the A sector, with a $SU(2)$ subgroup of the global symmetry group gauged. The commutant
$SO(2N_f)$ is the flavor group of the original theory. For $N=1$ the theory is $USp(2)\simeq SU(2)$ with 4 flavors, which has $SO(8)$ flavor
symmetry. In this case the A sector becomes free and describes four doublets of $SU(2)$. For $N=2$ the theory is $USp(4)$ with
six fields in the fundamental. This case has been studied by Seiberg and Argyres in \cite{AS}, where it was recognized that the A sector
coincides with the $E_7$ SCFT of Minahan and Nemeschansky \cite{MN} (so in this case there is an enhancement from the naive $SU(2)\times SO(12)$
to $E_7$). Indeed, the scale invariance of the curve requires (for $N=n-1$) that the $SU(2)$ beta function is zero, so the A sector
should give the same contribution as 4 flavors. This obviously works in the $N=1$ case and tells us that the $SU(2)$ central charge
is 8 in the other cases. This is precisely the value found in \cite{CDII} for the puncture with partition $\{2n-3,1,1,1\}$.

We can now determine the beta function of the $SU(2)$ gauge group emerging at the maximally singular point with the same technique adopted in the previous section: 
the curve for $\epsilon_A^2<\vert x\vert<\epsilon_{B}^{2}$ represents a tubular region associated with the $SU(2)$ gauge group. We can 
thus compute $a$ and $a_D$ and then determine the generalized coupling constant $\tau=\partial a_D/\partial a$.
\begin{equation*}
\begin{aligned}
 a=&\int_{\vert x\vert=\text{const.}}\lambda=2\pi i\alpha;\quad \alpha^{2}=-4\Lambda^{2N+2-2n}u_{N-n+2}-c_2,\\
 a_D=&\int_{\vert x\vert\sim\epsilon_{A}^2}^{\vert x\vert\sim\epsilon_{B}^2}\lambda=\alpha\left(\frac{2N-2n+2}{N-n+2}\log{\epsilon_{A}}+\text{const.}\right).
\end{aligned} 
\end{equation*}
Identifying as before $\epsilon_A$ with the energy scale we find
$$b_1=2\left(1-\frac{1}{N-n+2}\right),$$ which is the contribution to the beta function of the $D_{N-n+2}$ theory. Indeed, this is the
expected result, since the contribution from the A sector, as we have just seen, saturates the $SU(2)$ beta function.

The case $n=1$ deserves some comments: the A 
sector becomes trivial and we are left with the $D_{N+1}$ theory, which has $SU(2)$ flavor symmetry and not $SO(4)$! Let us analyze the curve
carefully in this case:
$$xy^2=(x^{N+1}-\dots-u_Nx+\tilde{c}_{2})(x^{N+1}-\dots-(u_N\pm4\Lambda^{2N})x+\tilde{c}_{2})-c_2x-\tilde{c}_2^2.$$
Scaling towards the small $x$ region we find $$y^2=\pm4\Lambda^{2N}(x^{N+1}-\dots-u_Nx)-c_2\pm4\Lambda^{2N}\tilde{c}_2,$$
where the casimir associated to the $SU(2)$ global symmetry is $c_2\pm4\Lambda^{2N}\tilde{c}_2$.
The $\pm$ term reflects the fact that there are two maximally singular vacua, corresponding to $u_N=\pm2\Lambda^{2N}$ (clearly,
this is true also for $n>1$).  The $n=1$ case is special because the two quadratic casimirs of $SO(4)$ enter symmetrically in the scaled
curve and in each one of the two singular points only an $SU(2)$ subgroup acts. Of course, it is well known that this occurs in
the $N=1$ case (i.e. the $USp(2)\simeq SU(2)$ gauge theory with two massless flavors): in this case the two singular points describe
two hypermultiplets which are neutral under a $SU(2)$ subgroup of the $SO(4)$ flavor symmetry group.

We are now in a position to determine the infrared physics at the Chebyshev point. The SW curve and differential are \cite{CKM}
$$y^2\approx x^{2n},\quad \lambda\approx\frac{y}{x^n}dx.$$ This are precisely the curve and differential at the maximally singular
point of $USp(2n)$ theory with $2n$ flavors, in which the B sector describes a doublet of $SU(2)$. We thus propose that the low-energy 
description at the Chebyshev point of $USp(2N)$ theory with $2n$ massless flavors includes:
\begin{itemize}
\item An abelian $U(1)^{N-n}$ sector, with massless particles charged under each $U(1)$ subgroup.
\item The A sector described above, with global symmetry $SU(2)\times SO(4n)$.
\item A third sector consisting of two hypermultiplets, whose symmetry is $SU(2)$. The gauging of the diagonal $SU(2)$ 
couples the last two sectors.
\end{itemize}
Notice that for $n=2$ the A sector becomes free and describes four doublets of $SU(2)$.

\section{$SO(N)$ SQCD}

We can extend this analysis to theories with gauge group $SO(N)$, that exhibit the same phenomena described
in the previous section (coalescence of r vacua and flavor symmetry enhancement in the massless limit). We analyze first theories
with $N$ even and then those with $N$ odd.

\subsection{$SO(2N)$ SQCD with $2n$ flavors}

Let us consider $SO(2N)$ gauge theory with $N_f=2n$ flavors. The theory becomes superconformal for $n=N-1$ and in the massless limit has $USp(4n)$
flavor symmetry. The SW curve and differential are
\begin{equation}\label{SO}
y^2=xP_{N}^{2}(x)-4\Lambda^{4N-4n-4}x^3\prod_i(x-m_{i}^{2}), 
\end{equation}
\begin{equation}\label{dif}
 \lambda=\frac{\sqrt{x}}{2\pi i}d\log\left(\frac{xP_{N}(x)-\sqrt{x}y}{xP_{N}(x)+\sqrt{x}y}\right),
\end{equation}
where $P_N(x)=x^N-\sum_{k=1}^{N-1}u_kx^{N-k}-(u_N)^2$ ($u_1$ is a Coulomb branch coordinate in this case as well). With usual manipulations we can rewrite the curve as
$$y^2=x(x^N-\dots-u_N^2)(x^{N}-\dots+(4\Lambda^{2N-2n-2}-u_{N-n-1})x^{n+1}-\dots-u_N^2)-\sum_{k=1}^{2n}c_{2k}x^{2n+3-k}.$$
Here we have simply redefined $u_{N-n-1}+2\Lambda^{2N-2n-2}\rightarrow u_{N-n-1}$. Turning off all the parameters we then get the maximally singular
point: \begin{equation}\label{max}
        y^2=x^{N+n+2}(x^{N-n-1}+4\Lambda^{2N-2n-2}),\quad\lambda=\frac{y}{x^{n+2}}dx.
       \end{equation}
Rescaling the curve as before we obtain ($t=y/x^{n+1}$)
\begin{equation}\label{socurve}
\begin{aligned}
t^2=&-\sum_{k=1}^{2n}c_{2k}x^{1-k}+\left(x^{N-n}-\dots-u_{N-n}-\dots-\frac{u_N^2}{x^{n}}\right)\\
&\times\left(x^{N-n-1}-\dots+(4\Lambda^{2N-2n-2}-u_{N-n-1})-\dots-\frac{u_N^2}{x^{n+1}}\right).
\end{aligned}
\end{equation}
We can now introduce the two sectors imposing $\vert x\vert\sim\epsilon_A^2$ and $\vert x\vert\sim\epsilon_B^2$. Setting $c_{2k}\sim O(\epsilon_{A}^{2k})$ leads to
$t\sim\epsilon_A$ in the A sector and $t\sim\epsilon_{B}^{N-n}$ in the second one, so we deduce $$\epsilon_A=\epsilon_{B}^{N-n}.$$ The same 
argument we gave in sections 2 and 3 then assigns $$u_i\sim O(\epsilon_{B}^{2i})\; i=1,\dots,N-n;\quad u_{N-n+i}\sim O(\epsilon_{A}^{2+2i})\; i<n;\quad u_N\sim O(\epsilon_{A}^{n+1}).$$ 

The SW curve in the B sector is the by now familiar curve for the $D_{N-n}$ theory:
$$t^2=4\Lambda^{2N-2n-2}(x^{N-n}-\dots-u_{N-n})-c_{2}.$$
In the conformal case $N=n+1$ this sector is trivial and describes a doublet of hypermultiplets when $N=n+2$.
The A sector is described by the curve
$$t^2=\left(u_{N-n}-\dots-\frac{u_{N}^{2}}{x^n}\right)\left(4\Lambda^{2N-2n-2}-\frac{u_{N-n}}{x}-\dots-\frac{u_{N}^{2}}{x^{n+1}}\right)-\sum_{k=1}^{2n}c_{2k}x^{1-k}.$$ 
This curve has $2n+2$ branch points and contrary to the $USp(2N)$ case the A sector is never free. The global symmetry group is $SU(2)\times USp(4n)$
and, as usual, the $SU(2)$ gauge group is gauged. In the scale invariant case we recover a S-dual description similar to the one for $USp(2N)$: the 
B sector is trivial and we are left with the A sector with a $SU(2)$ subgroup of the flavor symmetry group gauged. Now one can repeat the
calculation of the $SU(2)$ beta function with the same technique adopted in sections 2 and 3; the result is $$b_1=2\left(1-\frac{1}{N-n}\right).$$
This coincides again with the contribution of the $D_{N-n}$ theory, so the contribution from the A sector must saturate the $SU(2)$ beta
function as before.

\begin{figure}
\centering{\includegraphics[width=.3\textwidth]{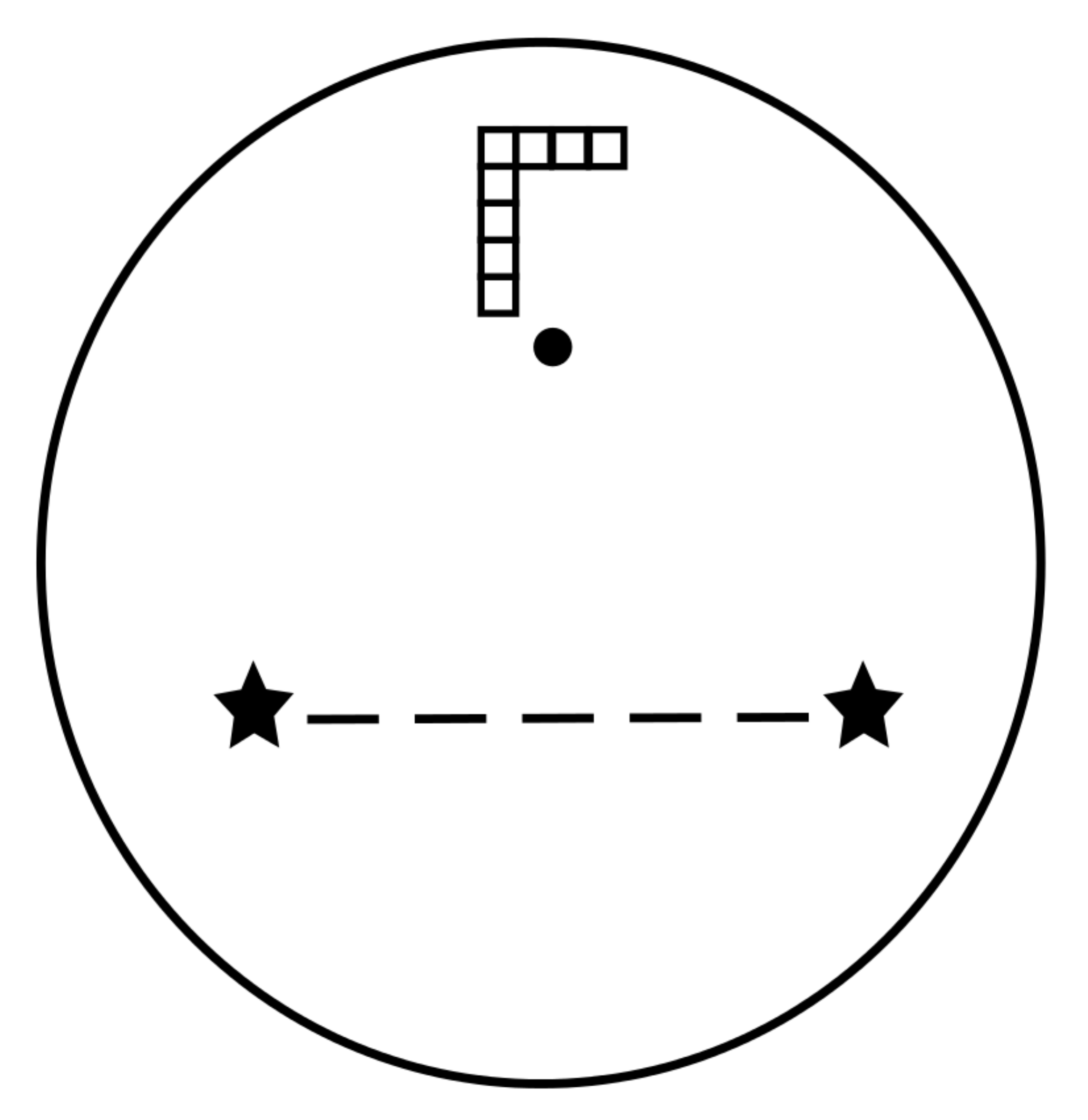}} 
\caption{\label{spdual}\emph{The three punctured sphere associated to the SCFT entering in the S dual description of $SO(2N)$ with $2N-2$ flavors (in this case $N=4$). 
The black dot indicates the D-partition and we have drawn the corresponding Young tableau. The two maximal C-partitions are indicated with $\star$.
To visualize the presence of the twist we draw a dashed line.}}
\end{figure}

Also in this case a description in terms of 6d $(2,0)$ $D_{n+1}$ theory compactified on a three punctured sphere is available (for $n>1$).
The only new ingredient is the presence of black puntures (in the notation of \cite{DT}), or C-partitions in the language of 
\cite{CDII,CDT}. To determine the theory, the simplest way is to notice that the A sector emerges in the dual description (of the 
strong coupling limit) of the scale invariant $SO(2N)$ SQCD. The collision of the simple punctures produces as before the 
D-partition (or grey puncture) described by Young tableau with $2n+2$ boxes, organized in a row of lenght four and the others of unit lenght (see figure\ref{spdual}). 
This puncture gives rise to a $SU(2)$ global symmetry group with central charge $k=8$, which is precisely the value needed to saturate the beta function. The 
remaining two punctures are described by a Young tableau with $2n$ boxes and a single row. The pole structure for 
the k-differentials encoded in this puncture has been determined in \cite{DT} and is $\{1,\dots,2n-1;n+1/2\}$ \footnote{The algorithm 
for determining the pole structure for general C-partitions is different with respect to the one described in section 3 and in this paper
we will not need it. The interested reader can find an exhaustive discussion on this point in \cite{CDT}.}. The fractional degree of the pole for 
$\tilde{\phi}_{n+1}$ is not a problem, since we have two such punctures. Turning around one of them we find $\tilde{\phi}_{n+1}\rightarrow-\tilde{\phi}_{n+1}$, 
which is precisely the action of the $\mathbb{Z}_{2}$ outer automorphism of the $D_{n+1}$ Lie algebra \cite{TII} (see figure\ref{spdual}). The k-differentials can thus be written as
$$\phi_{2k}=\frac{u_{2k}z}{(z-1)^2}\left(\frac{dz}{z}\right)^{2k}\; k=2,\dots,n;\quad\tilde{\phi}_{n+1}=\frac{u_{n+1}\sqrt{z}}{z-1}\left(\frac{dz}{z}\right)^{n+1}.$$
Using (\ref{sw}) we find the following SW curve: $$v^{2n+2}=\frac{z}{(z-1)^2}\left(\sum_{k=2}^{n}u_{2k}v^{2n+2-2k}+u_{n+1}^{2}\right).$$ With the same manipulations
described in section 3 we get precisely the curve for the A sector.

The case $n=1$ deserves some comments: turning off all the mass deformations the curve for the A sector becomes
$$t^2=-\frac{u_{N}^{2}}{x}\left(4\Lambda^{2N-4}-\frac{u_{N}^{2}}{x^{2}}\right).$$ It describes a rank one, scale invariant theory
with a Coulomb branch coordinate of dimension 2. Rank one scale invariant theories are indeed completely classified \cite{MNI} 
(see also \cite{AWIII}) and just from these data 
we can identify the A sector with the origin of the Coulomb branch of $SU(2)$ theory with 4 massless flavors.
This theory has central charges $a=23/24$, $c=7/6$ and $SO(8)$ global symmetry with central charge $k_{SO(8)}=4$ (see e.g. \cite{TIII}). 
Our A sector has instead $SU(2)\times USp(4)$ flavor symmetry. Here
we see once again the phenomenon first described by Argyres and Seiberg in \cite{AS}: $SO(8)$ has a maximal $SU(2)\times USp(4)$ subgroup and by
gauging the $SU(2)$ factor we recover the $USp(4)$ symmetry of the parent gauge theory. The $SU(2)$ central charge can be computed
using the formula given in \cite{AS} $$k_{SU(2)}=\mathcal{I}_{SU(2)\hookrightarrow SO(8)}k_{SO(8)},$$ where $\mathcal{I}$ is the 
embedding index. Using for example that the $\bf{8}_{V}$ of $SO(8)$ decomposes as $\bf{8}_{V}=(\bf{3},\bf{1})\oplus(\bf{1},\bf{5})$ under $SU(2)\times USp(4)$,
(indeed the result does not depend on the representation chosen) we obtain \cite{AS} $$\mathcal{I}_{SU(2)\hookrightarrow SO(8)}=\frac{T({\bf3})+5\cdot T({\bf1})}{T(\bf{8}_V)}=2.$$
We thus find that the $SU(2)$ central charge is 8, which is precisely the value needed to saturate the beta function.
As a final remark, summing the contribution to $a$ and $c$ coming from the $SU(2)$ gauge group and from the A
sector we get precisely the central charges for the $SO(4)$ theory with two flavors, which is nothing but the $SU(2)\times SU(2)$
gauge theory with two hypermultiplets in the $(\bf{2},\bf{2})$.

\subsection{$SO(2N+1)$ SQCD with $2n+1$ flavors}

The above analysis can be repeated for $SO(2N+1)$ gauge theories with odd number of flavors $N_f=2n+1$. The SW curve and differential
are \begin{equation}\label{BN}
y^2=xP_{N}^{2}(x)-4\Lambda^{4N-4n-4}x^2\prod_i(x-m_{i}^{2}), 
\end{equation}
\begin{equation}\label{swdif}
 \lambda=\frac{\sqrt{x}}{2\pi i}d\log\left(\frac{xP_{N}(x)-\sqrt{x}y}{xP_{N}(x)+\sqrt{x}y}\right),
\end{equation}
Redefining $u_{N-n-1}+2\Lambda^{2N-2n-2}\rightarrow u_{N-n-1}$ we can rewrite the curve as
$$y^2=-\sum_{i=1}^{2n+1}c_{2i}x^{2n+3-i}+x(x^{N}-\dots-u_N)(x^{N}-\dots+(4\Lambda^{2N-2n-2}-u_{N-n-1})x^{n+1}-\dots-u_N).$$
The most singular point can be found setting all $C_{2i}$ and $u_i$ to zero:
$$y^2=x^{N+n+2}(x^{N-n-1}+4\Lambda^{2N-2n-2});\quad\lambda\approx\frac{y}{x^{n+2}}dx.$$
As before, when $N=n+1$ the theory is conformal and this point coincides with the origin of the Coulomb branch. When $N=n+2$ we
recover the Chebyshev point, where the curve degenerates as $y^2\approx x^{N_f+3}$.

We can now set $t=y/x^{n+1}$ and introduce the A and B sectors, in which $\vert x\vert\simeq\epsilon_A^2$
and $\vert x\vert\simeq\epsilon_B^2$ respectively. The same argument given in the previous sections leads us to the relation $\epsilon_A=\epsilon_{B}^{N-n}$
and to the assignment $$u_i\sim O(\epsilon_{B}^{2i})\; i=1,\dots,N-n;\quad u_{N-n+i}\sim O(\epsilon_{A}^{2+2i}).$$ 
The curves describing the theories in the two sectors can now be readily identified: the B sector is the $D_{N-n}$ theory
(when $N=n+2$ it describes a doublet of $SU(2)$ and when $N=n+1$ becomes trivial) and the curve for the A sector is
\begin{equation}\label{eea}
t^2=-\left(u_{N-n}+\dots+\frac{u_N}{x^{n}}\right)\left(4\Lambda^{2N-2-2n}-\frac{u_{N-n}}{x}-\dots-\frac{u_N}{x^{n+1}}\right)-\sum_{i=1}^{2n+1}c_{2i}x^{1-i}. 
\end{equation}
One can easily see from the above curve that this sector has $SU(2)\times USp(4n+2)$ global symmetry. To identify the theory, let us start from the $n=1$ case. Turning off
the mass parameters we are left with the curve (setting $4\Lambda^{2N-4}=2$) $$t^2=-\frac{u}{x}\left(2-\frac{u}{x^2}\right).$$ 
To bring it to a more familiar form, it is now convenient to define $y=tx^2$. We then find $$y^2=-ux(2x^2-u);\quad\frac{\partial\lambda}
{\partial u}=\frac{dx}{y}.$$ Making then the change of variables $$y=-\frac{\tilde{y}}{2u};\quad x=-\frac{\tilde{x}}{2u},$$ we recognize the curve
for the $E_7$ SCFT of Minahan and Nemeschansky \cite{MN}: $$\tilde{y}^2=\tilde{x}^3-2u^3\tilde{x};\quad\frac{\partial\lambda}{\partial u}=\frac{d\tilde{x}}{\tilde{y}}.$$ 
The fact that only a $SU(2)\times USp(6)$ subgroup of $E_7$ appears in (\ref{eea}) has been explained in \cite{AW,AWII}: this is a submaximal mass deformation
of the $E_7$ theory and enters in the S-dual description of $SO(5)$ gauge theory with three flavors at the infinite coupling point, as shown in \cite{AW}.

We already encountered this theory in section 3: it is the A sector for $USp(2N)$ SQCD with six flavors. Comparing equations
(\ref{ea}) and (\ref{eea}), we can see that the analogy between the A sectors of these two theories is not limited to this case!
The A sectors of $SO(2N+1)$ SQCD with $N_f=2n+1$ flavors and $USp(2N)$ SQCD with $N_f=2n+4$ flavors are described by the same curve
(once we have set to zero the mass deformations), with Coulomb branch coordinates of the same scaling dimension. However, the 
flavor symmetry groups are different: $USp(4n+2)$ and $SO(4n+8)$ respectively. Based on the analysis of the $n=1$ case, it is
natural to suggest that the first theory represents a submaximal mass deformation of the second one. This is not surprising 
since the SW curves and differentials for $SO(2N+1)$ SQCD with $N_f=2N-1$ and $USp(2N)$ SQCD with $N_f=2N+2$ coincide in the
massless case.

\section{Conclusions}

We have made a systematic analysis of singular points in $\mathcal{N}=2$ SQCD with classical gauge groups, focusing on the 
maximally singular points in the moduli space. We have seen that, in order to satisfy the constraint on the scaling dimensions 
of mass parameters, we are forced to introduce different scale invariant sectors. The introduction of two sectors, which is the simplest
possibility, leads to a unique answer which is consistent with all the strong coupling dualities found recently and allows to 
satisfy the constraints imposed by superconformal invariance. 

We found a common structure for the low-energy description at these points, which is schematically given by:
\begin{itemize}
\item  An abelian sector.
\item The B sector, which is always given either by a $D_N$ theory (with $N>2$) or by a doublet of hypermultiplets. In both cases
the flavor symmetry is $SU(2)$.
\item The A sector, with (at least) $SU(2)\times G$ flavor symmetry, where $G$ is the flavor symmetry of the parent gauge theory. 
This is the only sector that changes as we vary the gauge group and in most cases admits a six-dimensional description.
\item An infrared free $SU(2)$ gauge multiplet coupled to sectors A and B.
\end{itemize}
At the maximally singular point the abelian sector just describes a number of decoupled vector multiplets, as pointed out in \cite{GS},
whereas at points which are not lifted by the $\mathcal{N}=1$ perturbation it includes massless hypermultiplets charged under
each $U(1)$ factor. Chebyshev points in $USp$ and $SO$ gauge theories fall in this second class and are characterized by a free B
sector, which describes two massless hypermultiplets. This is not the case for $SU(N)$ gauge theories. A particularly simple case is given by $USp$ theory with 4 flavors: in this
case the A sector becomes free and it turns out that the Chebyshev point admits a lagrangian description.

One possible future direction is to analyze more general theories and see whether different structures emerge in the infrared at
the maximally singular points. The Argyres-Douglas theories studied in \cite{XIE} certainly play an important role. In particular,
it would be interesting to see whether the constraints we imposed in this note always lead to a unique answer. 
Another interesting question is how to use our results to explore the properties of the $\mathcal{N}=1$ theories obtained adding
a superpotential for the chiral field in the adjoint.

\section* {Acknowledgments} It is a pleasure to thank my advisor Ken Konishi for many illuminating discussions at various stages
  of this project. I would also like to thank Alessandro Tanzini, Giulio Bonelli and Yuji Tachikawa for helpful comments.


\end{document}